\documentclass[%
 preprint, 
 amsmath,amssymb,
 aps, physrev,
]{revtex4-2}

\usepackage{graphicx}
\usepackage{dcolumn}
\usepackage{bm}
\usepackage{color}
\usepackage{booktabs} 

\newcommand{\beq}{\begin{equation}}
\newcommand{\eeq}{\end{equation}}

\begin{document}

\preprint{APS/123-QED}

\title{\textbf{First Principles Study of the Fermi Surface Topology of CeCu$_{2}$Si$_{2}$} 
}%

\author{Roxanne M. Tutchton}
\affiliation{Theoretical Division, Los Alamos National Laboratory, Los Alamos, New Mexico 87545, USA}
  \email{Contact author: rtutchton@lanl.gov}
  
\author{Jean-Pierre Julien}%
\affiliation{Université Grenoble Alpes, CNRS, Institut NEEL, F-38042 Grenoble, France}%

\author{Qimiao Si}
\affiliation{Department of Physics and Astronomy, Rice University, Houston, Texas 77005, USA}%

\author{Jian-Xin Zhu}
\affiliation{Theoretical Division, Los Alamos National Laboratory, Los Alamos, New Mexico 87545, USA}%
\affiliation{Center for Integrated Nanotechnologies, Los Alamos National Laboratory, Los Alamos, New Mexico 87545, USA}

\date{\today}

\begin{abstract}
Since the discovery of heavy-fermion superconductivity in CeCu$_{2}$Si$_{2}$, the material has attracted great interest particularly with regard to the nature of the superconducting pairing and its mechanism. Consequently, it is essential to better understand the electronic Fermi surface topology and its role in strong antiferromagnetic fluctuations. The standard density functional theory method is insufficient to model the interplay of strong onsite Coulomb repulsion in localized 4{\it f}-electrons and their hybridization with itinerant ligand-orbital electrons. We have performed electronic ground state calculations on CeCu$_{2}$Si$_{2}$ using the Gutzwiller wavefunction approximation. The Gutzwiller approximation captures the quasiparticle band renormalization from the strong onsite Coulomb repulsion. We have performed an analysis of this effect on the electronic structure and the Fermi surface topology by varying the interaction strength and taking into account the crystal-field splitting. Using the de Haas van Alphen effect, the extremal Fermi surface cross-sectional areas were calculated to quantify the effects of quasiparticle mass renormalization on the Fermi surface. Our results confirm the presence of two Fermi surface sheets corresponding to the heavy (488m$_{e}$) and light (4.35m$_{e}$) quasiparticles when the crystal-field splitting is accounted for on equal footing with the electronic correlations. This method gives the best agreement with experimental measurements as well as the renormalized band method. 
\end{abstract}

\maketitle

\section{Introduction}\label{intro}

The unconventional superconductor (SC), CeCu$_{2}$Si$_{2}$, has been a material of interest for research in both the SC community and the heavy fermion community for over forty-five years. In 1979 Steglich {\it et al}. published their findings on CeCu$_{2}$Si$_{2}$ showing a low temperature transition into a superconducting state at around 0.5 K \cite{1979Steglich}. This marked the first demonstration of superconducting pairing occurring in a metal driven by many-body interactions. It was postulated in this discovery work that the interactions would be ``probably magnetic in origin" arising from the low temperature anomalies characterized by an instability of the 4{\it f} shell. Subsequent studies on CeCu$_{2}$Si$_{2}$ would point to the importance of the Fermi surface and crystal field splitting in understanding the nature of the {\it f} electron behavior \cite{1981Horn, 1985Steglich, 2006Fulde}.

In 1990, magnetic oscillations were observed for CeCu$_{2}$Si$_{2}$, and the de Haas Van Alphen (dHvA) effect was measured for one of the quasiparticle sheets making up  the Fermi surface. They approximated quasiparticles with an effective mass of $\sim$4.5-6 times the mass of an electron\cite{1990Hunt}. The measurement of the Fermi surface through dHvA effects is one of most useful ways to understand the {\it f} electrons directly. Therefore, shortly after these measurements were published, Hirama and Yanase presented density function theory (DFT) calculations in the local density approximation (LDA) \cite{1991Harima, 1992Harima}, which predicted a Fermi surface comprised of three sheets. Though they were unable to conclusively determine how their calculations compared with experiments at the time, their work was foundational to the understanding of electronic states in CeCu$_{2}$Si$_{2}$. A year later, Zwicknagl {\it et al}.\cite{1993Zwicknagl} published an electronic structure study using renormalized band theory. This method predicted two sheets in the Fermi surface, one with light quasiparticles and the other with heavy quasiparticles. The light quasiparticle sheet was in good agreement with the previous LDA calculations, but the heavy sheet was found to be highly sensitive to the renormalizing effects of strongly correlated electrons. Their study also suggested that the superconducting phase transition is likely driven by a Fermi surface topological transition in the heavy quasiparticle system.

Many of these early predictions started to be confirmed with improved sample preparation techniques and theoretical methods \cite{1993Goremychkin, 1984Batlogg, 2003Yuan, 2007Ning, 2009Matsumoto,2010Arndt}. In 2011, neutron scattering experiments identified antiferromagnetic excitations as the driving force for superconducting pairing in CeCu$_{2}$Si$_{2}$ \cite{2010Stockert}. Unlike conventional SCs where superconductivity is destroyed by magnetic ions, CeCu$_{2}$Si$_{2}$ relies on a periodic dense array of magnetic 
moments from the Ce$^{3+}$ ions' $f$ electrons to transition to a superconducting state \cite{2012Steglich}. This complex and unique interplay between the magnetic ordering and SC has been explored in numerous experimental and theoretical studies \cite{2010Stockert,2011Vieyra,2004Stockert,2008Eremin,2020Li}. Given the confirmed importance of the $f$ electrons in the physics of the system, in 2014, dynamical mean field theory (DMFT) was applied to CeCu$_{2}$Si$_{2}$ to attempt a better explanation of the heavy fermion physics of the material and explore the importance of the crystal field splitting  \cite{2014Pourovskii}. This method predicted a single heavy fermion sheet in the Fermi surface with a quasiparticle effective mass of ~200m$_{e}$. Zwicknagl {\it et al}.\cite{2016Zwicknagl} once again approached the material with renormalized band theory in 2016, predicting a Fermi surface comprised of two sheets, with a light quasiparticle effective mass of ~5m$_{e}$ and a heavy quasiparticle mass of 500m$_{e}$. 

Several theoretical studies have confirmed the importance of the electron correlations and crystal field effects in CeCu$_{2}$Si$_{2}$\cite{1993Goremychkin, 2007Ning, 2015Ikeda, 2021Song, 2020Li, 2021Nica}, including the work of Amorese {\it et al.} in 2020\cite{2020Amorese}, which does a through study of the crystal field states using renormalized band theory and X-ray absorption spectroscopy (XAS). Subsequently, Li {\it et al}.\cite{2018Li} applied LDA+U within the random phase approximation (RPA)\cite{2008Eremin} to predict two Fermi surface sheets in good agreement with Zwicknagl {\it et al}.\cite{1993Zwicknagl, 2016Zwicknagl}, and Li {\it et al.}\cite{2020Li} and Luo {\it et al.}\cite{2020Luo} applied DMFT to the electronic analysis and temperature dependence respectively. Additional works have been done exploring the quasiparticle behavior and impurity effects using DFT+$U$\cite{2022Zhao, 2024Zhao}. The existence of the heavy quasiparticle Fermi surface sheet has been experimentally observed as of 2021 in Wu {\it et al}.\cite{2021Wu} using angle resolved photo emission spectroscopy (ARPES), which qualitatively confirms the presence of the two Fermi surface sheets calculated by both renormalized band theory and LDA+U methods, though in this case their calculations predicted the mass renormalization of the heavy band to be $\sim120m_{e}$.

Here we present an alternative theoretical approach to calculate the Fermi surface topology and mass enhancement for CeCu$_{2}$Si$_{2}$ using the Gutzwiller Wavefunction Approximation (GWA)\cite{1965Gutzwiller}. Through this investigation we show that the GWA is capable of reproducing the two Fermi Sheets indicated by experiment and that crystal field splitting (CFS) is as significant to the underlying physics of the electronic structure of CeCu$_{2}$Si$_{2}$ as the electronic correlations and spin-orbit coupling (SOC), which we demonstrate through comparisons to GWA calculations that do not take CFS into account. We have also determined the sign of the charge and crystal field parameter, $\alpha$, to be in agreement with the X-ray scattering measurements of Willers {\it et al.}\cite{2012Willers} and Rueff {\it et al.}\cite{2015Rueff}. Furthermore, the increased computational efficiency of the GWA method allows us to perform calculations on complex systems, such as CeCu$_{2}$Si$_{2}$, with less computational expense than DMFT \cite{2015Lanata,2020Tutchton,2020Chiu}. In the following sections, we describe the theoretical framework for addressing the CFS within the GWA (section~\ref{theory}), the results of our electronic structure calculations and analysis (section~\ref{secElecStruct}), and our analysis of the Fermi surface topology (section~\ref{secFermiSurfTop}). We compare our results for calculations performed with and without electronic correlations and with and without CFS incorporated into the correlation effects. A conclusion is given in section~\ref{conclusion}. The details of our methodology are described in section~\ref{methods}. A full account of our analysis of the Coulomb parameter tests is given in the Sumplementary Material\cite{SupMat}. 

\section{Results and Analysis}\label{results}
\subsection{Structure and Crystal Field}\label{theory}
\begin{figure*}[htb]
\begin{center}
\includegraphics[width=0.7\textwidth]{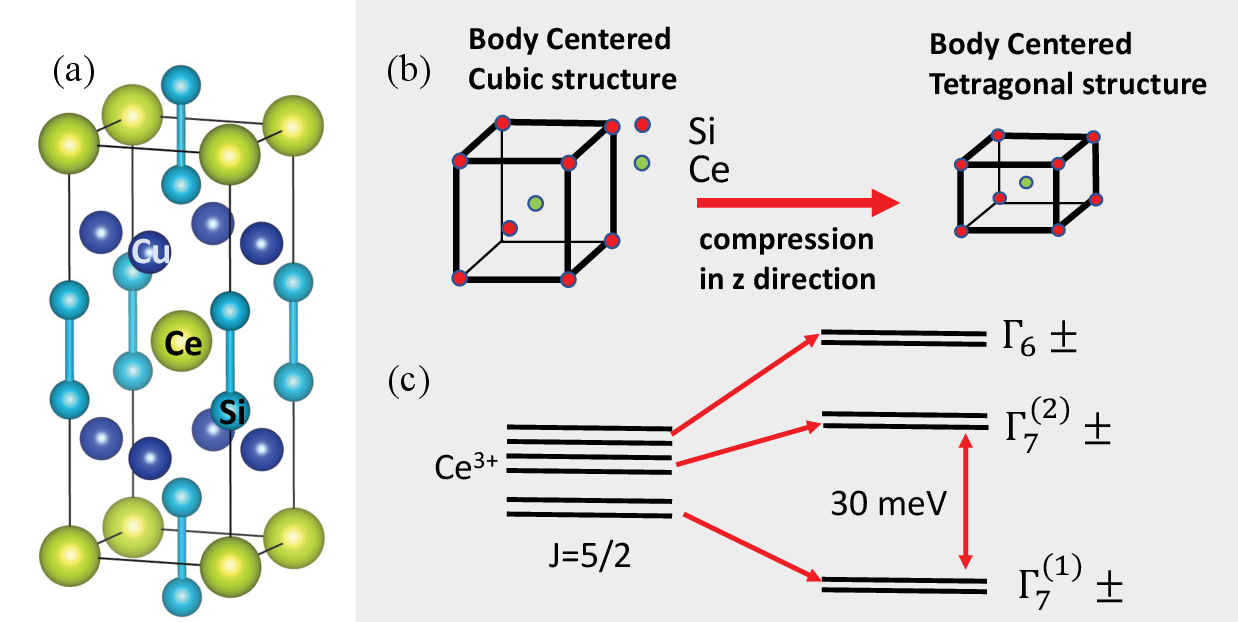}
\end{center}
\caption{ {\bf Crystal Structure of CeCu$_{2}$Si$_{2}$} where (a) shows the ideal crystal configuration with the yellow atoms representing Ce, the dark blue representing Cu, and the light blue representing Si. The mechanism of the crystal field splitting is shown in (b). Here the red atoms represent Si and the green represent Ce. The diagram shows the compression of the z-axis which results in the body centered tetragonal (BCT) configuration of the Si and Ce atoms centered in the real crystal structure shown in (a). This results in the fine structure splitting depicted in (c) with the energetic separation of $\Gamma_{7}^{(1)}$ and $\Gamma_{7}^{(2)}$.} 
\label{CFS_struct}
\end{figure*}

The atomic environment of the CeCu$_{2}$Si$_{2}$ structure and crystal field splitting is shown in Fig. ~\ref{CFS_struct} with the Ce atom in a body centered tetragonal (BCT) environment of 8 Si atoms and 8 Cu atoms. From the GWA calculations (see Sec.~\ref{methods}) we have determined the electronic environment and used a model to determine the CFS parameters for the electronic ground state. Our results are consistent with those given by the group theory~\cite{1963Condon}, and a full derivation of those parameters is given in Sec. I.A of the Supplementary Materials~\cite{SupMat}. To summarize, the $j=5/2$ ground state of the Ce atom can be split into three Kramers doublets, two $\Gamma_{7}^{(1,2)}$ and a $\Gamma_{6}$ as shown in FIG.~\ref{CFS_struct}(c). Thus the ground state doublet can be written in the $j m_j$ basis as

\begin{align}
|\Gamma_{7}^{(1)}\pm \rangle= \alpha \; | \frac{5}{2},\pm\frac{5}{2} \rangle + \sqrt{1-\alpha^2}\; | \frac{5}{2},\mp\frac{3}{2} \rangle. \label{Gamma7_1}
\end{align}
Then the doublet
 \begin{align}
 |\Gamma_{7}^{(2)}\pm \rangle=  \sqrt{1-\alpha^2}\;\; | \frac{5}{2},\pm\frac{5}{2} \rangle - \alpha\; | \frac{5}{2},\mp\frac{3}{2} \rangle \label{Gamma7_2}
\end{align}
 gives the first excited states, with the energy gap between this doublet and the ground state doublet, experimentally known to be of the order of 30 meV\cite{1993Goremychkin}.  Finally, the highest energy states constitute the doublet
\begin{align}
|\Gamma_{6}\pm \rangle= | \frac{5}{2}, \pm \frac{1}{2} \rangle. \label{Gamma6}
\end{align}

\noindent The sign of $\alpha$ determines the orientation of the the angular distribution of the $\Gamma_{7}$ state. For $\alpha<0$ the angular distribution is along the [011] plane, as opposed to $\alpha>0$, which would orient the distribution on the [100] plane\cite{2012Willers, 2020Amorese}. The measured magnitude of $\alpha$ determined from neutron diffraction is $\sim|0.88|$\cite{1993Goremychkin}. X-ray diffraction measurements place the value closer to $\sim|0.62|$\cite{2020Amorese}. In our validation model we have used $\alpha=-0.88$, which gives us the 30 meV crystal field splitting between the ground state, $\Gamma_{7}^{(1)}$, and first excited state, $\Gamma_{7}^{(2)}$. This is used to inform our analysis of the CFS effects on the electronic structure of CeCu$_{2}$Si$_{2}$ calculated using the implementation in the CyGutz code as detailed in Lanat\`a {\it et al}\cite{2017Lanata}. This is in good qualitative agreement with the X-ray absorption spectroscopy and band renormalization theory analysis in Amorese {\it et al.} \cite{2020Amorese} as well as the neutron-scattering investigations of Goremychkin, \textit{et. al.}~\cite{1993Goremychkin}. For more information on the calculation parameters and interpretation of the result, see the Supplementary Materials \cite{SupMat} Secs. I.A, III and IV.A.2.

\subsection{Electronic Structure}\label{secElecStruct}
\begin{figure*}[htb]
\begin{center}
\includegraphics[width=0.9\textwidth]{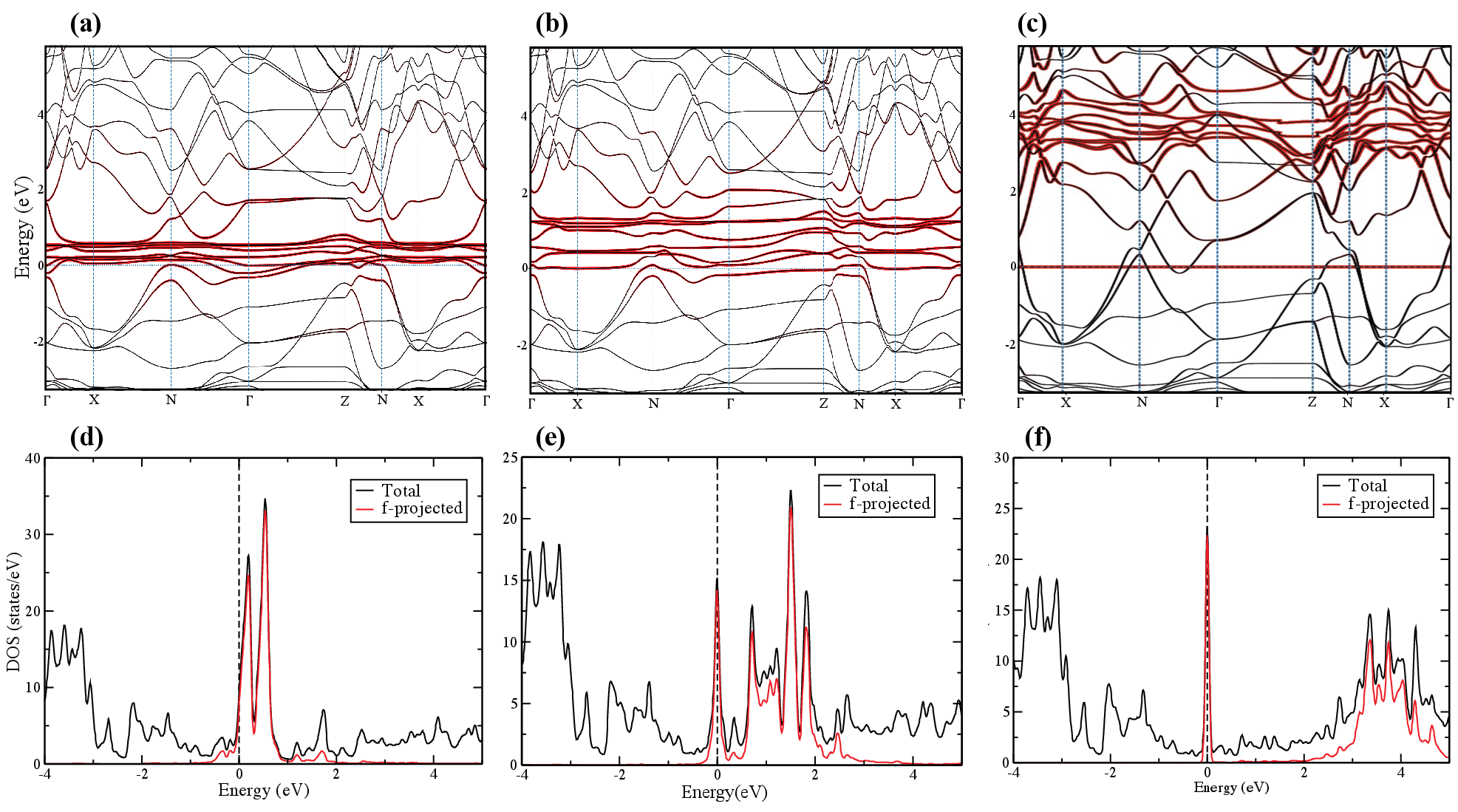}
\end{center}
\caption{ {\bf The electronic ground state calculations} for CeCu$_{2}$Si$_{2}$ calculated using the GWA in the generalized gradient approximation (GGA) with spin orbit coupling. (a) and (d) give the electronic band structure and DOS respectively for the $U=0.0$ case, (b) and (e) show the $U=5.1$ eV calculation with CFS accounted for in the GWA band renormalization, and (c) and (f) show the results for $U=6.0$ eV with CFS. The black lines in the band structures show the eigenenergy bands, and red lines indicate the {\it f}-electron contribution. Similarly, in the DOS, the black line is the total DOS and the red line is the projected {\it f}-electron DOS. The high symmetry path through the BZ is shown in FIG.S2 of the Supplementary Materials\cite{SupMat}.} 
\label{elect_struct}
\end{figure*}

The electronic band dispersion, electron density of states (DOS), and Fermi surface for CeCu$_{2}$Si$_{2}$ were calculated within the GWA with and without electronic correlations. The calculations use a DFT basis as described in section~\ref{methods}. Since DFT includes CFS, all of the initial calculations take the crystal field effects into account at the weakly correlated level. To better understand the interplay between the electron correlation and the CFS, we have also tested the GWA with and without CFS included in the band renormalization parts of the calculation\cite{2017Lanata}. The results without correlations ($U=0$) or additional CFS are given in FIG.\ref{elect_struct} (a) and (d). The three corresponding Fermi surface sheets are given in FIG.\ref{FStop} in order of descending band number from top to bottom for each doubly degenerate band. These results are in close agreement with those calculated in the early 1990's by Harima {\it et al}\cite{1991Harima, 1992Harima}. The results with electronic correlations and CFS are shown in FIG.\ref{elect_struct} (b) and (e) with a Coulomb interaction strength of $U=5.1$ eV and in (c) and (f) with $U=6$ eV. An exchange coupling parameter of $J=0.7$ eV was used for each calculation where $U>0$ and relativistic (SOC) effects are considered in a paramagnetic configuration for each case (see section \ref{compmethods}).

The localization of the {\it f}-electrons can be seen both in the band dispersion and the DOS. For the case without electron-electron correlations or GWA CFS, the two spikes in the DOS just above the Fermi energy are the $j=5/2$ and $j=7/2$ orbital sub-shells, respectively (FIG.\ref{elect_struct}d). This indicates that the orbital character at the Fermi energy is dominated by the six-fold degenerate $j=5/2$ state of the 4{\it f}-electron. When electronic correlations ($U=5.1$ eV) are taken into account, the CFS effect is enhanced. As such, the DOS has a sharp peak at the Fermi energy and a more disperse arrangement of the higher energy quasiparticle states (FIG.\ref{elect_struct}e), which is consistent with the lifting of the Ce $j=5/2$ electrons due to CFS (FIG.~\ref{CFS_struct}). This appears in the band dispersion (FIG.\ref{elect_struct}b) as a broadening of the {\it f}-electron occupation and a flattening of the bands around the Fermi energy leading to two, rather than three, sheets in the Fermi surface (FIG.\ref{FStop}b). Calculations for $U=5.1$ eV and $U=6$ eV are shown in FIG.\ref{elect_struct} to demonstrate the sensitivity of the electronic ground state to the presence and strength of the electron-electron interactions. The effects of CFS can be more clearly seen in the analysis of the Fermi surface in FIG.\ref{FStop}, which is discussed in section~\ref{secFermiSurfTop}. 

As the electron correlations become stronger in the system, the $j=5/2$ peak grows sharper and narrower, becoming more delta-like. Conversely, the $j=7/2$ occupation appears at a higher energy while becoming more disperse as $U$ increases to 6 eV. This is consistent with expected SOC splitting in a strongly coupled system. A full analysis of the Coulomb parameter strength was performed for cases $U=0$ to $U=6$eV and is discussed in more detail in section III.A of the Supplementary Material\cite{SupMat}, but the impact of the electronic correlations are most dramatic between $U=5$ eV and $U=6$ eV where the system itself begins to have precarious stability eventually leading to a collapse of the bands around the Fermi energy once the correlation strength exceeds a critical point between $5\;\text{eV}<U<6$ eV. For our particular method, we found a critical interaction strength of $U=5.1$ eV gave the best overall convergence and agreement with past studies \cite{1993Zwicknagl,2016Zwicknagl,2020Amorese,2021Wu,2021Song,2024Zhao}. 

Notably, the sensitivity of the electronic structure is even more pronounced in the presence of CFS as shown by the eigenvalues of the 4{\it f} quasiparticle matrix, $Z$, given in Table~\ref{Z_data} along with the corresponding orbital occupations, $n$. $Z$ is the matrix of quasiparticle weights derived from the "Gutzwiller self-energy" defined by equation 14 of Lanat\`a {\it et al.}\cite{2017Lanata} (and discussed in section III.A.1 of the Supplementary Material\cite{SupMat}) to describe the rotationally invariant slave boson (RISB) mean-field theory\cite{2015Lanata, 2017Lanata}. The eigenvalues of this matrix can be related to the effective mass of the quasi particle as $1/Z=m_{qp}^{*}/m_{e}$. These eigenvalues have been calculated with and without CFS effects and recorded in Table~\ref{Z_data}, which compares these quasiparticle weights in each case for increasing values of $U$. In the case where CFS is neglected, the {\it f} electrons are split into $j=5/2$ and $j=7/2$ sub-orbitals, and the magnitude of $Z$ decreases as $U$ increases indicating a gradual increase in the effective quasiparticle mass. The orbital occupation increases very slightly with increasing $U$ in the $j=5/2$ subshell and decreases with increasing $U$ in the $j=7/2$ subshell. For the calculations including CFS, the $j=5/2$ subshell is split into the $\Gamma_{7}^{(1)}$, $\Gamma_{7}^{(2)}$, and $\Gamma_{6}$ states (FIG.~\ref{CFS_struct}). In the $\Gamma_{7}^{(1)}$ Kramers doublet, defined by equation~\ref{Gamma7_1}, $Z$ decreases sharply between $U=5.1$ eV and $U=6$ eV.  Here we see that this slight increase of the Coulomb interaction strength corresponds to a relatively sharp decrease in the quasiparticle weight in $\Gamma_{7}^{(1)}$ and corresponding increase in the effective quasiparticle mass, $m_{qp}$. This is consistent with the evolution of the DOS and our exploration of the Fermi surface topology (see section~\ref{secFermiSurfTop}), which indicates an increasingly sensitive and coupled relationship between the electronic correlations and the CFS between $U=5$ eV and 6 eV. Indeed, at $U=6$ eV, $Z$ approaches zero. The total orbital occupation is consistent with a localized 4$f^{1}$ configuration with the greatest occupancy in the $\Gamma_{7}^{(1)}$ doublet. This is in agreement with XPS\cite{1980Lasser} and XAS\cite{1987Edwards} observations as well as previous theoretical analysis based on DMFT calculations\cite{2020Li}. An account of the complete results for the quasiparticle weights and orbital occupations from $U=0$ eV to 6 eV is given in the Supplementary Material\cite{SupMat}.

\begin{table}[]
\caption{{\bf Eigenvalues of the 4{\it f} quasiparticle matrix, $Z$, and corresponding orbital occupations, $n$}, at $U$ values 0.0\footnote{The quasipatricle weight for U=0 (as in a regular DFT calculation) is Z=1.0. The 0.96 value is an artifact from the valence truncation applied by the GWA calculations.}, 5.1, and 6.0 eV. The top half of the table displays the results that include CFS effects in the GWA and the bottom part of the table show results that neglect the CFS in the GWA.\label{Z_data}}
\begin{center}
\setlength{\tabcolsep}{8pt}
\begin{tabular}{@{}ccccccc@{}}
\toprule\hline 
\multicolumn{1}{l}{With CFS}    &\multicolumn{2}{c}{$\Gamma_{7}^{(1)}$} & \multicolumn{2}{c}{$\Gamma_{7}^{(2)}$}  & \multicolumn{2}{c}{$\Gamma_{6}$} \\ \midrule 
U (eV)   & Z      & n        & Z      & n     & Z      & n                     \\ \hline
\multicolumn{1}{l|}{0.0}   & 0.96 & \multicolumn{1}{l|}{0.48} & 0.96 & \multicolumn{1}{l|}{0.02} & 0.96 & 0.04 \\
\multicolumn{1}{l|}{5.1} & 0.47 & \multicolumn{1}{l|}{0.50} & 0.96 & \multicolumn{1}{l|}{0.01} & 0.99 & 0.01                \\
\multicolumn{1}{l|}{6.0}   & 0.00 & \multicolumn{1}{l|}{0.50} & 0.97 & \multicolumn{1}{l|}{0.01} & 0.96 & 0.01  \\ 

\midrule \hline
\multicolumn{1}{l}{Without CFS}   &\multicolumn{3}{c}{$5/2$} & \multicolumn{3}{c}{$7/2$}  \\ \midrule 
U (eV)     & Z      & n    &    & Z      & n    &                    \\ \hline
\multicolumn{1}{l|}{0.0}     & 0.96 & \multicolumn{2}{l|}{0.10} & 0.96 & \multicolumn{2}{l}{0.04}  \\ 
\multicolumn{1}{l|}{5.1}     & 0.48 & \multicolumn{2}{l|}{0.14} & 0.74 & \multicolumn{2}{l}{0.02}  \\
\multicolumn{1}{l|}{6.0}     & 0.32 & \multicolumn{2}{l|}{0.15} & 0.75 & \multicolumn{2}{l}{0.01}  \\  
\bottomrule \hline \hline
\end{tabular}
\end{center}
\end{table}

\subsection{Fermi Surface Topology}\label{secFermiSurfTop}

Analysis of the Fermi surface topology gives a clear indication of the impact of the CFS in the presence of strong electron correlations. There are numerous theoretical studies and experimental measurements that look at the Fermi surface of CeCu$_{2}$Si$_{2}$. Based on the combined efforts of magnetic oscillation studies \cite{1990Hunt} and ARPES experiments\cite{2021Wu}, it has been experimentally confirmed that there are two sheets in the Fermi surface. Theoretical methodologies from the LDA\cite{1991Harima, 1992Harima} to LDA+$U$ in the random phase approximation (RPA)\cite{2008Eremin,2015Ikeda, 2021Song, 2021Wu, 2024Zhao} to renormalized band theory\cite{1993Zwicknagl,2016Zwicknagl} and DMFT\cite{2014Pourovskii} have been applied to understand the Fermi surface topology. The theoretical results appear to depend heavily on the method, choice of electronic correlation strength, and the presence of CFS. 

\begin{figure*}[htb]
\begin{center}
\includegraphics[width=0.9\textwidth]{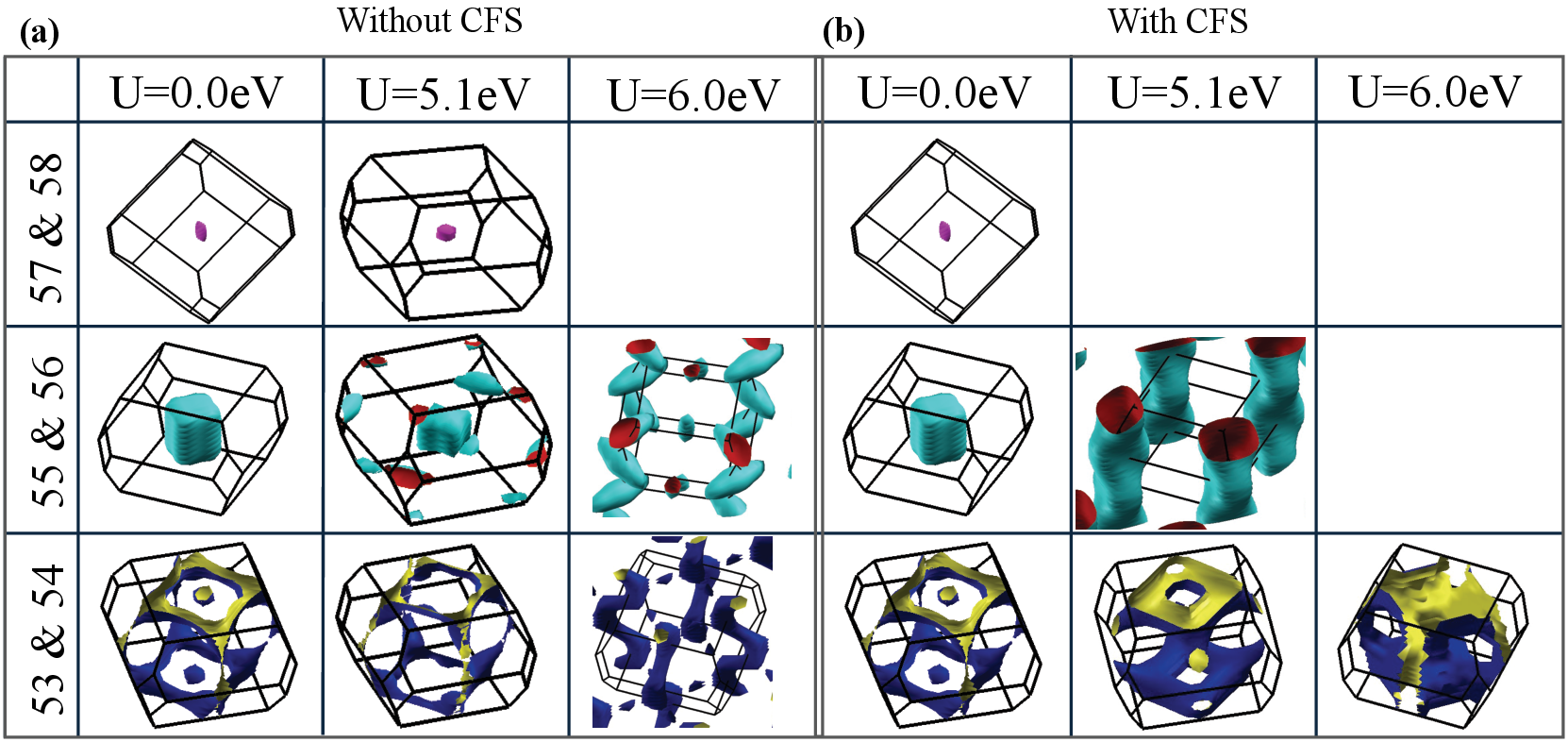}
\end{center}
\caption{ {\bf The Fermi surface topology} for CeCu$_{2}$Si$_{2}$ calculated using the GWA both without (a) and with (b) CFS. The Fermi sheets are are shown by descending band number and energy. Each sheet is doubly degenerate. Calculation for $U=0.0$, $U=5.1$ eV, and $U=6.0$ eV are shown. The complete results for $U=0$ though 6 eV are discussed in the Supplementary Materials\cite{SupMat}.} 
\label{FStop}
\end{figure*}

In the early works of Harima {\it et al.}\cite{1991Harima,1992Harima}, the LDA method predicted three Fermi surface sheets. Our GGA calculations at $U=0$, agree closely with both the qualitative character of the Fermi sheet topologies as well as the dHvA simulation results for the maximal frequencies and cyclotron masses. In our survey of the Hubbard parameter strength in the GWA\cite{SupMat}, we have tracked the evolution of the Fermi surface topology (FIG.~\ref{FStop}), and simulated the dHvA extremal frequencies for an external {\bf B} field parallel to the $z$-axis. The dHvA results for the Fermi surface obtained using the simulation method of Rourke and Julian\cite{2012Rourke} described in section~\ref{dhvasim} are given in Table~\ref{dHvA_data}, which shows the results when CFS is included and when it is neglected. For $U=0$, the impact of CFS on the Fermi surface topology is negligible. 

It is well documented that better agreement with the current experimental consensus can be achieved in the Fermi Surface by including strong electron-electron correlations (via a Coulomb interaction). The DFT+$U$+RPA type methods tend to over estimate the impact of the 4{\it f} electrons in the conduction bands, but they are capable of giving results that compare qualitatively well with experiment. These methods also appear to be successful with smaller $U$ values according to Ikeda {\it et al.}\cite{2015Ikeda}. In our calculations using a GWA approach without CFS the correlation strength needs to be large in order to predict a Fermi surface containing two sheets, rather than three. Even at $U=6$ eV this method does not give the expected cylindrical heavy band predicted by the DFT+$U$+RPA and renormalized band theory methods (FIG.~\ref{FStop}). Additionally, we explored a DMFT approach (neglecting CFS) discussed in the Supplementary Materials\cite{SupMat}. These Fermi surface topology results compare nearly exactly to our GWA calculation at the same value of $U$, which is shown in FIG.S10. There have been several DMFT studies of CeCu$_{2}$Si$_{2}$\cite{2014Pourovskii, 2020Li, 2020Luo}. The study by Pourovskii {\it {\it et al.}.}\cite{2014Pourovskii} included CFS and predicted a single sheet in the Fermi surface, which is at odds with previous DFT+$U$ methods and experimental measurement. It is, however in close agreement with our GWA+CFS results at $U=6$ eV (FIG.~\ref{FStop}).

The electronic structure results from section~\ref{secElecStruct} are an indication that the CFS effects are significant to the band dispersion and therefor the Fermi surface. This is demonstrated by the comparison of FIG.\ref{FStop} (a) and (b), where (b) shows the evolution of the Fermi surface topology with increasing $U$ with CFS included. At $U=5.1$ eV with CFS, there are two sheets making up the Fermi surface and the dHvA data is given in Table~\ref{dHvA_data}. The qualitative Fermi surface is in close agreement with the renormalized band theory studies of Zwicknagl {\it et al.}\cite{1993Zwicknagl, 2016Zwicknagl}, as well as the DFT+$U$+RPA methods\cite{2015Ikeda, 2021Song}. The effective cyclotron mass (eq.~\ref{dHvA_mass}) calculated with CFS also agrees closely with the renormalized band theory calculations which predicted an effective mass of $~5m_{e}$ for the light quasiparticle sheet and $~500m_{e}$ for the heavy sheet. Our calculations (with CFS included) predict 4.35$m_{e}$ and 488$m_{e}$ for the light and heavy sheets respectively (see section III.B for the Supplementary Material\cite{SupMat}). These results show that CFS needs to be treated within the Gutzwiller mean-field theory for correlations to achieve results consistent with experimental observations.

\begin{table*}
\caption{{\bf The dHvA and volume data for Fermi surface calculations with and without CFS in the GWA renormalization.} Frequencies are given in kilotesla (kT), and corresponding effective masses are in units of electron mass ($m_{e}$). Band numbers are given for the odd bands in each degenerate pair. Reciprocal occupied (electron bands 55-58) and unoccupied (hole bands 53 and 54) Fermi surface volumes are given in units of {\AA}$^{-3}$. The electron occupation, e$^{-}$, is the approximate number of electrons in the band. All measurements were taken for a magnetic field parallel to the $z$ axis (${\bf B}\parallel z$). The Hubbard parameter strengths at $U=0.0$, 5.1, and 6.0 eV are compered. \label{dHvA_data}}
\begin{center}
\setlength{\tabcolsep}{9pt}
\renewcommand{\arraystretch}{0.9}
\begin{tabular}{@{}cc|cccc|cccc@{}}
\toprule \hline
      &      &  \multicolumn{4}{c}{with CFS} & \multicolumn{4}{c}{without CFS} \\ 
      \hline
U (eV) & band & $f$   & $m^{*}$ & $V_{FS}$& e$^{-}$ &  $f$     & $m^{*}$  & $V_{FS}$  & e$^{-}$  \\
\hline
0.0 & 53 & 10.49 & 3.282    & 0.239 & 1.840 & 10.49 & 3.282  & 0.242 & 1.838 \\ 
    & 55 & 3.397  & 2.480    & 0.210 & 0.141 & 3.397  & 2.480  & 0.215 & 0.144 \\
    & 57 & 0.093  & 0.884    & 0.002 & 0.001 & 0.093  & 0.884  & 0.002 & 0.001 \\
\hline
5.1 & 53 & 11.45  & 4.353    & 0.518 & 1.653 & 10.35 & 9.214  & 0.203 & 1.864 \\
    & 55 & 5.910  & 488.0    & 1.007 & 0.676 & 2.741  & 72.230 & 0.149 & 0.100 \\
    & 57 &    -   &  -       &  -    &   -   & 0.357  & 3.419  & 0.005 & 0.003 \\
\hline
6.0 & 53 & 18.346 & 2000     & 2.127 & 1.427 & 10.85 & 15.908 & 0.232 & 1.844 \\
    & 55 &   -    &   -      &   -   &    -  & 3.333  & 66.60 & 0.153 & 0.103 \\ 
\hline
\bottomrule

\end{tabular}
\end{center}
\end{table*}
\section{Conclusion}\label{conclusion}
In the last $~30$ years there have been a number of studies using various methods to describe the Fermi surface of CeCu$_{2}$Si$_{2}$. Experimental measurements have confirmed the existence of a Fermi Surface comprised of two sheets\cite{1990Hunt, 2021Wu}, several theoretical and experimental studies have explored the importance of the CFS and electron correlation effects on the overall electronic structure and the superconducting pairing\cite{1993Goremychkin, 2007Ning, 2012Rourke, 2012Willers, 2015Ikeda, 2021Song, 2020Li, 2020Amorese}, and new advances have made it possible to perform more accurate and detailed analysis of the intricate coupling and correlation effects in this complex material. 

In this study we have presented a systematic analysis of the electronic structure and Fermi surface topology of CeCu$_{2}$Si$_{2}$ as a function of the Coulomb interaction strength with a focus on the CFS effect. We have performed calculations using the GWA to show that the resulting Fermi surface topology is highly sensitive to methodology, electron-electron interactions, and CFS treatment. De Haas van Alphen simulations were applied to the Fermi surfaces to analyze the Fermi sheets. Our GWA results calculated using a Hubbard parameter of $U=5.1$eV and including CFS in the mean-field theory are found to be in agreement with both experimental measurements as well as the renomalized band theory. More broadly, our results suggest that the CFS plays an important role in driving heavy fermion behavior in the presence of strong correlations in the electronic structure of CeCu$_{2}$Si$_{2}$. The qualitative and quantitative aspect of the Fermi surface are also extremely sensitive to small changes in the electron correlation, particularly in the presence of CFS. Based on the analysis of the electronic DOS, this is due to the complexity of the 4{\it f} electron behavior present in the heavy Fermi sheet.

Our results highlight the necessity for a thorough treatment of the electronic coupling in order to better understand the nature of the heavy-fermion superconductivity and formation of Cooper pairs. Future studies of the temperature dependence (which has been touched on in Luo {\it {\it et al.}.}\cite{2020Luo}) and possible electron-phonon coupling contribution would aid in forming an even more comprehensive picture of CeCu$_{2}$Si$_{2}$. 

\section{Methods}\label{methods}
\subsection{Computational Methods}\label{compmethods}
To account for the strong electronic coupling in CeCu$_{2}$Si$_{2}$ we combine standard density functional theory with the GWA method implemented in the CyGutz code \cite{2015Lanata, 2017Lanata}. This method uses the full-potential, linearly augmented plane wave (FPLAPW) DFT of WIEN2k \cite{wien2k} as its basis and implements a combination of the GWA and the rotationally invariant slave boson (RISB) method to account for strong electronic correlations. Relativistic spin-orbit coupling (SOC) effects were included with a k-point grid of $12\times 12 \times 12$ and muffin tin radii of 2.50$a_{0}$ for the Ce and Cu atoms and 1.93$a_{0}$ for the Si atoms, where $a_{0}$ is the Bohr radius. The cut-off parameter was set to $R_{mt}\times K_{max}=9.0$. A generalized gradient approximation (GGA-PBE)\cite{1996Perdew} exchange correlation functional was used throughout all calculations. A double counting correction is done in the fully-localized-limit with the local orbital occupation updated in the outer electron density self-consistent loop, and the Coulomb interaction strength was tested at values from $U=$0 eV to $U=$6.0 eV\cite{SupMat}. We tested the GWA band renormalization implementation with and without CFS effects included beyond the initial DFT. For all calculations where $U>0$, Hund's exchange coupling parameter, $J$, was set at 0.7 eV. For comparison, a GGA+DMFT calculation was performed for $U=6$ eV and $J=0.7$ eV. Results and details of the methodology are given in the Supplementary Materials, sections SII and SIII\cite{SupMat}. 

\subsection{de Haas van Alphen simulations}\label{dhvasim}
Analysis of the Fermi surfaces was done using numerical calculations of the dHvA effect as implemented by Rourke and Julian\cite{2012Rourke}. By applying a magnetic field to the system, oscillations in the magnetic susceptibility can be determined from the changes in the number of occupied Landau levels as a function of the reciprocal magnetic field, $1/{\bf B}$\cite{2012Rourke, 2015Jiao}. Then the dHvA frequency can be expressed as\\
\begin{equation}
f_{i} = \frac{1}{\Delta (1/\mathbf{B})} = \frac{\hbar}{2\pi e} A_{i}
\label{dHvA_freq}
\end{equation}
where $e$ is the elementary charge of an electron, and $A_{i}$ is the extremal cross-sectional area of the $i^{th}$ branch of the Fermi surface in a plane perpendicular to $\mathbf{B}$. The effective carrier mass averaged around the extremal cyclotron orbits is also determined from
\begin{equation}
m^{*} = \frac{\hbar^{2}}{2\pi e}\left. \frac{\partial A}{\partial E}\right\rvert_{E=E_{F}}
\label{dHvA_mass}
\end{equation}
where $m^{*}$ is in units of the electron mass, $m_{e}$. This gives the instantaneous effective mass of the quasiparticles in the Fermi sheet at a particular orbit rather than the overall effective mass of the quasiparticles in the energy band. The results of the dHvA analysis  are given in Table~\ref{dHvA_data}.

\bibliography{Ce122_library}
\section*{Acknowledgements}

The authors thank Chao Cao, Bogyu Jang, Yongxin Yao, E. M. Nica, F. Steglich and G. Zwicknagl for useful discussions.
This work was carried out under the auspices of the U.S. Department of Energy (DOE) National Nuclear Security Administration under Contract No. 89233218CNA000001. The Fermi-surface topology analysis work was partly supported by the LANL LDRD Program and the U.S. Department of Energy, Office of Science, National Quantum Information Science Research Centers, Quantum Science Center. The GWA simulation work was supported by the NNSA Advanced Simulation and Computing Program. It was, in part, supported by the Center for Integrated Nanotechnologies, a DOE BES user facility, in partnership with the LANL Institutional Computing Program for computational resource as well as NERSC.The work at Rice has primarily been supported by the National Science Foundation under Grant No. DMR-2220603, by the Robert A. Welch Foundation Grant No. C-1411 and the Vannevar Bush Faculty Fellowship ONR-VB N00014-23-1-2870. Q. S. acknowledges the hospitality and support by a Ulam Scholarship from the Center for Nonlinear Studies at Los Alamos National Laboratory during the initial stage of the work.

\section*{Author contributions statement}

R.M.T performed calculations and prepared the results and analysis. J-P.J formulated the theory for the Crystal Field analysis and advised on GWA theory. Q.S formulated the initial motivation for the work and advised on conceptualization for the project. J-X.Z formulated the initial motivation for the project and advised at all levels. All authors participated in the discussion of results and reviewed the manuscript.

\section*{Additional information}

The authors declare that they have no competing financial interests.
\end{document}